\documentclass[12pt]{article}
\usepackage{epsfig}
\usepackage{amssymb}

\input{epsf.sty}

\newcommand{\la}[1]{\label{#1}}
 
\newcommand{\be}{\begin{equation}}
\newcommand{\ee}{\end{equation}}
\newcommand{\ba}{\begin{eqnarray}}
\newcommand{\ea}{\end{eqnarray}}
\newcommand{\bi}{\begin{itemize}}
\newcommand{\ei}{\end{itemize}}

\newcommand{\tr}{{\rm Tr\,}}

\newcommand{\trq}{{\rm \hat{T}r\,}}
\newcommand{\ex}{{\rm e}}

\newcommand{\nn}{\nonumber}

\newcommand{\bfx}{{\bf x}}
\newcommand{\bfy}{{\bf y}}

\newcommand{\hu}{\hat{U}}
\newcommand{\hp}{\hat{P}}
\newcommand{\hl}{\hat{L}}
\newcommand{\htm}{\hat{T}}
\newcommand{\op}{{\cal O}}

\newcommand{\RR}{{\rm I\kern -.2em  R}} 
\newcommand{\eq}{Eq.~}
\newcommand{\eqs}{Eqs.~}
\newcommand{\fig}{Fig.~}

\newcommand{\se}{Sec.~}

\newcommand{\rmF}{{\rm F}}

\newcommand{\rmA}{{\rm A}}
\newcommand{\rmav}{{\rm av}}

\def\lsi{\raise0.3ex\hbox{$<$\kern-0.75em\raise-1.1ex\hbox{$\sim$}}}
\def\gsi{\raise0.3ex\hbox{$>$\kern-0.75em\raise-1.1ex\hbox{$\sim$}}}

\begin{document}
 
\begin{titlepage}
\begin{flushright}
MIT-CTP-3510
\end{flushright}
\begin{centering}
\vfill
 
{\bf The Polyakov Loop and its Relation to Static Quark Potentials\\ and Free Energies}

\vspace{0.8cm}
 
Oliver Jahn$^1$ and Owe Philipsen$^2$

\vspace{0.3cm}
{\em $^{\rm 1}$
Center for Theoretical Physics, Massachusetts Institute of Technology,\\
Cambridge, MA 02139-4307, USA\\}
{\em $^{\rm 2}$
Dept.~of Physics and Astronomy, University of Sussex, Brighton BN1 9QH, UK}

\vspace*{0.7cm}
 
\begin{abstract}
  It appears well accepted in the literature that the correlator of
  Polyakov loops in a finite temperature system decays with the
  ``average'' free energy of the static quark-antiquark system, and
  can be decomposed into singlet and adjoint (or octet for QCD)
  contributions.  By fixing a gauge respecting the transfer matrix,
  attempts have been made to extract those contributions separately.
  In this paper we point out that the ``average'' and ``adjoint'' channels
  of Polyakov loop correlators are misconceptions.  We show analytically
  that {\it all} channels receive contributions from singlet states
  only, and give a corrected definition of the singlet free energy. 
  We verify this finding by simulations of the 3d SU(2) pure gauge
  theory in the zero temperature limit, which allows to cleanly
  extract the ground state exponents and the non-trivial matrix
  elements. The latter account for the difference between the channels
  observed in previous simulations.

\end{abstract}
\end{centering}
 
\noindent
\vfill
\noindent
 

\vfill

\end{titlepage}
 

\section{Introduction}

In recent years there has been a growing interest in assessing possible
contributions of color octet bound states to physical processes. While 
such states carry color charge and hence do not exist as asymptotic 
particle states, they may appear as virtual intermediate states in certain processes
to whose cross sections and branching ratios they should then measurably contribute.
Examples are quarkonium production where such contributions
arise naturally in the framework of Non-Relativistic QCD \cite{nrqcd}, 
and the QCD plasma phase in which colored states might be excited thermally \cite{sz}.
It is expedient to first try and understand such physics in the 
heavy quark limit, where bound states can be obtained from a non-relativistic 
Schr\"odinger equation. This requires knowledge of the non-perturbative static 
quark-antiquark potential as input, which may be extracted from lattice 
simulations of Wilson or Polyakov loops. 

While the Wilson loop probes the color singlet potential,
there seems to be a common understanding in the literature that the
correlator of Polyakov loops decays exponentially with the ``color average'' potential, which
decomposes into two parts: one related
to the color-singlet ($\rm 1$) potential and one to the color-octet or adjoint ($\rmA$)
potential \cite{bro,ls,nad}.
These potentials are thought to exist, and be distinct in the
zero-temperature limit, corresponding to different sectors in the Hilbert space
with color sources coupled to the 
singlet and adjoint
representations, respectively.  This is corroborated by perturbation theory
where the zero-temperature potentials can be computed to low orders in the
coupling and are found to be gauge-invariant.  To leading order one has \cite{bro,nad}
\be
(N^2-1) V_\rmA(r) = - V_1(r) + \op (g^4).
\ee

More recently it was shown that, by dressing (or gauge fixing) 
the Polyakov loop in a way that respects the transfer matrix,
it is possible to extract the singlet potential $V_1(r)$ from 
Polyakov loop correlators \cite{op1}, which suggests to obtain the adjoint channel by
subtracting the singlet piece from 
the average. This approach was used in several numerical investigations
of the different color channels at finite temperature \cite{lit},
for which the potentials turn into superpositions of Boltzmann weighted excitations and are interpreted
as free energies.
A recent simulation employing different gauges confirmed gauge independence 
for the zero temperature potentials, but observed gauge dependence 
at finite temperatures \cite{bbm}.

In this paper we clarify the situation regarding Polyakov loops.
Converting Euclidean expectation values into traces over states in 
the Hilbert space, we show analytically that Polyakov loop correlators
in {\it all} channels receive contributions from singlet states exclusively. 
Thus none of the channels or their combinations 
can serve to define an octet potential. 
Moreover, at non-zero temperature the standard 
Polyakov loop correlator does not average over the 
color channels, but probes a thermal sum of only singlet excitations.

We verify these statements by numerical simulations of the pure SU(2)
gauge theory in 2+1 dimensions in the zero temperature limit.
We compute the correlators in all channels at a series of low
temperatures and extract the energy of the lowest state as well as its weight in
the thermodynamic sum (its matrix element).  The energies are identical
within errors for all channels.  We further show that the putative
difference between the color channels, and in particular the 
repulsive r-dependence of the ``adjoint'' potential observed in previous 
simulations, is actually due to
the r-dependence of the matrix element of the singlet ground state.

\section{Operators for the different channels}

On an $L^3\times N_t$ lattice with periodic boundary conditions, 
we consider the untraced Polyakov loop (timeslices are labeled from $0...N_t-1$)
\be
L(\bfx)=\prod_{t=0}^{N_t-1}U_0(\bfx,t).
\ee
According to \cite{bro,ls,nad}, the color average potential at temperature
$T=1/(aN_t)$ is defined by
\be \label{vav}
\ex^{-V_{\rmav}(r,T)/T}=
\frac{1}{N^2}\langle \tr L^\dag(\bfx) \tr L(\bfy)\rangle 
=
\frac{1}{N^2}\ex^{-V_1(r,T)/T}+\frac{N^2-1}{N^2}\ex^{-V_{\rmA}(r,T)/T}.
\ee
The singlet and adjoint channels into which it decomposes follow after
projection on the corresponding representation matrices to be \cite{nad}
\ba
\ex^{-V_1(r,T)/T}&=&
\frac{1}{N}\langle \tr L^\dag(\bfx)L(\bfy)\rangle,\nn\\
\ex^{-V_{\rmA}(r,T)/T}
&=&\frac{1}{N^2-1}\langle \tr L^\dag(\bfx) \tr L(\bfy)\rangle
-\frac{1}{N(N^2-1)}\langle \tr L^\dag(\bfx) L(\bfy)\rangle.
\label{potdef}
\ea
The limit $N_t\rightarrow \infty$ takes us to zero
temperature, at which $V_{\rmav}, V_1, V_{\rmA}$ represent the ground state potentials. 
At finite $N_t$ viz.~$T$, they are instead Boltzmann weighted sums
over the excitation spectrum and interpreted as free energies.
Of the above expressions only the average correlator is gauge invariant.
In \cite{op1} it was shown that one can complete the singlet correlator gauge invariantly 
by dressing the Polyakov lines with some functional 
of spatial links, $\Omega[U_i]$, 
which under local gauge transformations transforms as
\be \label{otrafo}
\Omega^g(x)=g(x)\Omega(x)h^\dag (t).
\ee
Here $h(t)$ is an undetermined $SU(N)$ matrix which may be different in every timeslice.
Replacing $L(\bfx)$ in the above expressions by
\be
\tilde{L}(\bfx)\equiv \Omega^\dag (\bfx,0)L(\bfx)\Omega(\bfx,0),
\la{dress}
\ee
renders the singlet and adjoint correlators gauge invariant as well. 
Of course, $\Omega(x)$ may be interpreted as a gauge fixing function which is local in time
(such as e.g.~Coulomb gauge), in
which case the dressed correlators are equivalent to the original correlators in
a fixed gauge. We stress however that this does not affect the exponential decay. 
As was shown in \cite{op1,fp}, the Wilson loop and 
the correlator of gauge fixed temporal 
Wilson lines decay with the spectrum of the same transfer matrix, only the matrix 
elements are different. One can use this to re-express \eqs (\ref{potdef}) by the 
manifestly gauge invariant periodic Wilson loop. Choosing the gauge fixing functions in 
$\langle \tr \tilde{L}^\dag(\bfx)\tilde{L}(\bfy)\rangle$ 
to represent the spatial Wilson line (or ``string'') between $\bfx$ and $\bfy$, 
$\Omega(\bfx,t)\Omega^\dag(\bfy,t)=U(\bfx,\bfy;t)$, one obtains the periodic Wilson loop,
\be
\langle \tr \tilde{L}^\dag(\bfx)\tilde{L}(\bfy)\rangle = 
\langle \tr L^\dag(\bfx)U(\bfx,\bfy;0)L(\bfy)U^\dag(\bfx,\bfy;N_t)\rangle
=W(|\bfx-\bfy|,N_t),
\ee
which is equivalent to the singlet correlator \eq (\ref{potdef}) in axial gauge,
$U(\bfx,\bfy)=1$. This operator is cheap to compute, 
manifestly gauge invariant and the
one we shall use in our numerical investigation. 
However, our observations are valid 
for any other choice of $\Omega(x)$ which is local in time and transforms as \eq (\ref{otrafo}),
as well.

The original ``average'', ``singlet'' and ``adjoint/octet'' labeling of the correlators
refers to the transformation properties of the operators when
explicit fields for the static sources are introduced. At a given time,
interpolating operators for a static meson in a color singlet and adjoint state  
are, e.g.,
\be
O(\bfx,\bfy)= \bar{\psi}(\bfx)U(\bfx,\bfy)\psi(\bfy),\quad 
O^a(\bfx,\bfy)=\bar{\psi}(\bfx)U(\bfx,\bfx_0)T^aU(\bfx_0,\bfy)\psi(\bfy),
\la{ops}
\ee
with group generators $T^a$.
Here $\bfx_0$ represents the center of mass coordinate of the meson,
at which these operators transform as a singlet and adjoint under gauge transformations, respectively.
Integrating over
the static fields and using the completeness relation for $T^aT^a$, 
one finds for the corresponding correlation functions
\ba
\langle O(\bfx,\bfy;0)O^{\dag}(\bfx,\bfy;N_t)\rangle&\propto&
\langle \tr L^\dag(\bfx)U(\bfx,\bfy;0)L(\bfy)U^\dag(\bfx,\bfy;N_t)\rangle\nn\\
\langle O^a(\bfx,\bfy;0)O^{a\dag}(\bfx,\bfy;N_t)\rangle&\propto&
\left[\frac{1}{N^2-1}\langle \tr L^\dag(\bfx) \tr L(\bfy)\rangle\right.\\
& & \left.
-\frac{1}{N(N^2-1)}\langle \tr L^\dag(\bfx)U(\bfx,\bfy;0)
L(\bfy)U^\dag(\bfx,\bfy;N_t)\rangle\right] ,\nn
\ea
i.e.~\eqs (\ref{potdef}) in axial gauge. 

Expressing the Euclidean expectation values as Hamiltonian 
traces over complete sets of states
by means of the transfer matrix formalism \cite{cre,pos},
we find
\ba
\langle O(\bfx,\bfy;0)O^{\dag}(\bfx,\bfy;N_t)\rangle&\propto&
Z^{-1}\trq\left[\hu^\dag_{\alpha\beta}(\bfx,\bfy)(\htm^{N_t})_{\beta\gamma \delta\alpha}
\hu_{\gamma\delta}(\bfx,\bfy)\right],\nn\\
\langle O^a(\bfx,\bfy;0)O^{a\dag}(\bfx,\bfy;N_t)\rangle&\propto&
Z^{-1}\trq\left[\hu^{\dag a}_{\alpha\beta}(\bfx,\bfy)(\htm^{N_t})_{\beta\gamma \delta\alpha}
\hu^{a}_{\gamma\delta}(\bfx,\bfy)\right ] ,
\la{ham}
\ea
where $\hu^a(\bfx,\bfy)\equiv \hu(\bfx,\bfx_0)T^a\hu(\bfx_0,\bfy)$, and the transfer matrix is specified in
\eqs (\ref{tm},\ref{P33}) below.
Note that after integrating out the source fields, the operators 
in \eq (\ref{ops}) reduce to their pure gauge parts,
\be
O_{\alpha\beta}(\bfx,\bfy)=U_{\alpha\beta}(\bfx,\bfy),\quad
O^a_{\alpha\beta}(\bfx,\bfy)=U^a_{\alpha\beta}(\bfx,\bfy).
\la{ops1}
\ee 
In addition to transforming at $\bfx_0$ under the local gauge group, 
they also transform like a fundamental ($\rmF$)/antifundamental ($\bar{\rmF}$) representation at $\bfx,\bfy$, respectively.
When acting on the gauge-invariant vacuum, they hence generate states
living in the $\rmF(\bfx)\otimes \bar{\rmF}(\bfy)\otimes
\bf{1}(\bfx_0)=\rmF(\bfx)\otimes \bar{\rmF}(\bfy)$ and
$\rmF(\bfx)\otimes \bar{\rmF}(\bfy)\otimes \rmA(\bfx_0)$ sectors of
the full Hilbert space, respectively.
Keeping the sources saturating the indices at $\bfx,\bfy$ in mind,
we shall refer to these as the sectors of singlet and adjoint states.

Evidently, the exponential decay of both correlators 
in \eqs (\ref{ham}) is governed by the spectrum of the transfer matrix, 
\be \la{matel}
\langle \ldots \rangle \propto \sum_n \,c_n \ex^{-E_n/T},
\ee
and the energy levels are 
independent of the 
detailed form of the operators $U(\bfx,\bfy)$ (or other gauge 
fixing functions $\Omega(\bfx)$) \cite{op1}.
Based on this it was concluded that  
separate potentials in the adjoint/singlet channels can be obtained,
which then combine to the average potential as in \eqs (\ref{vav},\ref{potdef}).

However, this conclusion does not hold. 
As we shall now show, the so-called adjoint and average correlators are mislabeled: 
they receive contributions exclusively from singlet states.
Even though the operators in the adjoint correlator transform as an adjoint at $\bfx_0$,
the eigenstates contributing to it do not.
The crucial observation is that, after integrating out the source fields,
the transfer matrix in \eq (\ref{ham}) has acquired indices needed
for the time evolution of the string operators.
This reflects the fact that on all timeslices the Gauss law with static sources is
imposed
rather than the Gauss law without sources as in the case of
correlators of particle states.  
It turns out that, for both correlators, the states propagated by the transfer matrix
are in the singlet sector $\rmF(\bfx)\otimes \bar{\rmF}(\bfy)$. 
These states combine with the operators to give gauge-invariant, non-trivial matrix elements.

\section{Projection on the Hilbert space sectors}

This is best seen by using projection operators on the Hilbert space.
The transfer matrix in \eq (\ref{ham}) can be decomposed as
\be\label{tm}
(\htm)_{\alpha\beta\mu\nu}=\htm_0\hp_{\alpha\beta\mu\nu}^{\rmF\otimes\bar{\rmF}},
\ee where $\htm_0$ is the familiar transfer matrix in temporal gauge
acting on the Hilbert space in the presence of external charges
\cite{cre,pos}, i.e.~$\htm_0=\exp(-a\hat{H_0})$ with $\hat{H_0}$ the
Kogut-Susskind Hamiltonian \cite{ks}.  
The operator
\be \label{P33}
\hp_{\alpha\beta\mu\nu}^{\rmF\otimes\bar{\rmF}}= \int Dg \;
g^\dag_{\alpha\beta}(\bfx)g_{\mu\nu}(\bfy) \hat{R}[g], 
\ee
where $\hat{R}[g]$ imposes a gauge transformation $g(\bfx)$ on wave
functions, projects on the sector with a static quark-antiquark pair.
More specifically, it annihilates all wave functions not transforming
as $\psi_{\beta\mu}[U^g]=g(\bfx)_{\beta\gamma}
\psi_{\gamma\delta}[U]g^\dag_{\delta\mu}(\bfy)$ and maps
$\psi_{\beta\mu}$ to $\psi_{\nu\alpha}$ (see
Appendix~\ref{projectors}).  Note that this is the transformation
behavior of our singlet (at $\bfx_0$) wave functions. 
On the other hand, the operator projecting on
adjoint states is given by \be \la{adp}
\hp_{\alpha\beta\mu\nu a b}^{\rmF\otimes\bar{\rmF}\otimes \rmA}= \int
Dg \; g^\dag_{\alpha\beta}(\bfx)g_{\mu\nu}(\bfy)
D^{\rmA}_{ab}(g(\bfx_0))\hat{R}[g], 
\ee
 where $D^{\rmA}_{a b}(g)$ are
the representation matrices of the adjoint representation, 
\be
D^{\rmA}_{a b}(g)=2\,\tr(g^\dag T^agT^b).  
\ee

It is well known \cite{lw} that the ``average'' correlator can be expressed as
a quantum mechanical trace over a complete set of eigenstates of $T_0$,
\ba \label{avdec}
\frac{1}{N^2}\langle \tr L^\dag(\bfx) \tr L(\bfy)\rangle& =&
\frac{1}{N^2Z}\sum_{\alpha\beta}\trq[\htm_0^{N_t}\;\hp_{\alpha\alpha\beta\beta}^{\rmF\otimes\bar{\rmF}}]
=\frac{1}{N^4Z}\trq[\htm_0^{N_t}\;\hp^{\rmF\otimes\bar{\rmF}}]\nn\\ 
&=&\frac{1}{N^4}\sum_{n\alpha\beta}|\langle n_{\alpha\beta}|n_{\beta\alpha}\rangle|^2\; \ex^{-E_n/T} 
=\frac{1}{N^2}\sum_{n}\; \ex^{-E_n/T},
\ea
where we have used that $\hp^{\rmF\otimes\bar{\rmF}}\equiv
N^2\hp_{\alpha\alpha\beta\beta}^{\rmF\otimes\bar{\rmF}}$ is a projector.
The presence of this projector enforces that only singlet eigenstates 
$|n_{\alpha\beta}\rangle$ transforming
like $\psi_{\alpha\beta}$ above contribute to the correlator.
Hence all energy levels contributing to the thermodynamic sum are 
energies of singlet states, 
with no matrix elements besides the constant $1/N^2$.
To understand how this fits together with the supposed decomposition into
singlet and adjoint contributions, we similarly rewrite \eqs (\ref{ham}). 
Using periodicity of the trace, one finds
\ba
\langle O(\bfx,\bfy;0)O^{\dag}(\bfx,\bfy;N_t)\rangle&\propto&
Z^{-1}\,\trq[\htm_0^{N_t}\;\hu_{\gamma\delta}(\bfx,\bfy)\hu^\dag_{\alpha\beta}(\bfx,\bfy)
\, \hp_{\beta\gamma\delta\alpha}^{\rmF\otimes\bar{\rmF}}]\nn\\
&=&\frac{1}{N^2}\sum_{n}|\langle n_{\delta\gamma}|\hu_{\gamma\delta}(\bfx,\bfy)\hu^{\dag}_{\alpha\beta}(\bfx,\bfy)
|n_{\beta\alpha}\rangle |^2\; \ex^{-E_n/T}\nn\\
\langle O^a(\bfx,\bfy;0)O^{\dag a}(\bfx,\bfy;N_t)\rangle&\propto&
Z^{-1}\,\trq[\htm_0^{N_t}\;\hu^{a}_{\gamma\delta}(\bfx,\bfy)\hu^{\dag a}_{\alpha\beta}(\bfx,\bfy)
\,\hp_{\beta\gamma\delta\alpha}^{\rmF\otimes\bar{\rmF}}]\nn\\
&=&\frac{1}{N^2}\sum_{n}|\langle n_{\delta\gamma}|\hu^{a}_{\gamma\delta}(\bfx,\bfy)
\hu^{\dag a}_{\alpha\beta}(\bfx,\bfy)
|n_{\beta\alpha}\rangle |^2\; \ex^{-E_n/T}. \nn\\
\ea
These expressions reveal two non-trivial features:
firstly, even in the ``adjoint''
correlator projection is onto the
$\rmF(\bfx)\otimes \bar{\rmF}(\bfy)\otimes \bf{1}(\bfx_0)$ sector of Hilbert 
space, while the adjoint projector, \eq (\ref{adp}), does not appear at all.
Hence, the energy levels $E_n$ contributing to {\it both} expressions are the singlet ground state
potential and its excitations.
Second, the indices of the operators are
contracted with those of the eigenfunctions of the Hamiltonian, rather than among
each other. This means that there are non-trivial matrix elements depending 
on the operators
$U(\bfx,\bfy)$ (or other gauge fixing functions 
$\Omega(\bfx)$). As we shall see in \se \ref{numerics}, these
matrix elements are responsible for the structure observed in previous simulations 
in the literature \cite{op1,lit}.

\section{Spatial exponential decay}

The correlators \eqs (\ref{vav},\ref{potdef}) can of course equally be viewed as correlation functions in space, and often one is interested in their spatial decay with the separation $r=|\bfx-\bfy|$
between the static sources. In this case one performs an analogous analysis by defining a 
{\it spatial} transfer matrix, and corresponding Hamiltonian, which propagate states along
the axis defined by the charges. 
In this case one finds for the average correlator the well known result
\ba \label{avsp}
\frac{1}{N^2}\langle \tr L^\dag(\bfx) \tr L(\bfy)\rangle&=&
Z^{-1}\trq[\htm^{L-r}\;\tr \hl^\dag(\bfx_\perp)\,\htm^r \,\tr \hl(\bfy_\perp)]\nn\\
&\stackrel{L\rightarrow \infty}{\longrightarrow} &
\frac{1}{N^2}\sum_{n}|\langle 0|\tr \hl^\dag(\bfx_\perp)|n\rangle|^2\; \ex^{-rE_n}\, ,
\ea
where $\bfx_\perp$ stands for the coordinates perpendicular to the axis 
$\bfx-\bfy$.

The transfer matrix $\htm=\htm_0\hp$ acts on the sector of gauge invariant eigenstates of the 
spatial Hamiltonian. The eigenvalues $E_n$ depend on the finite $N_t$. 
In particular, for large $N_t$ the theory is
confining, and $E_n\sim N_t$ are energies of torelonic states, while for small $N_t$ they are
gauge invariant screening masses \cite{spat}.

For the singlet correlator some care has to be taken, 
because in general a gauge fixing function $\Omega(x)$ is non-local in the correlation direction
and thus prohibits the definition of a positive transfer matrix. However, our choice of axial gauge
is simply the spatial equivalent of temporal gauge, for which a positive transfer matrix exists and the
analysis can be performed. We find
\ba \label{singsp}
\frac{1}{N^2}\langle \tr L^\dag(\bfx)U(\bfx,\bfy;0) L(\bfy)U^\dag(\bfx,\bfy;N_t)\rangle& =&
Z^{-1}\trq[\htm^{L-r}\hl^\dag_{\alpha\beta}(\bfx_\perp)\htm_0^r\hl_{\beta\alpha}(\bfy_\perp)]\\
& =&
Z^{-1}\trq[\htm^{L-r}\hl^\dag_{\gamma\delta}(\bfx_\perp)\htm_0^r \hp_{\delta\beta\alpha\gamma}^{\rmF\otimes\bar{\rmF}} \hl_{\beta\alpha}(\bfy_\perp)]\nn\\
&\stackrel{L\rightarrow \infty}{\longrightarrow} &
\frac{1}{N^2}\sum_{n}|\langle n_{\alpha\beta}|\hl_{\beta\alpha}(\bfx_\perp)|0\rangle|^2\; \ex^{-rE_n}.\nn
\ea
In the second equality, we have used the transformation behavior
(\ref{transformation}) of $U$, applied to the Polyakov line $L$, which
is now transverse to the direction in which the
transfer matrix acts.  Note that the energy eigenvalues in this case are those of
the transfer matrix $\htm_0$, describing the sector with charges
propagating spatially. Hence, with respect to the spatial decay 
there is a difference between the two channels of correlators.
This difference is also reflected in perturbation theory,
where to leading order \eq (\ref{avsp}) is dominated by two gluon exchange,
compared to one gluon exchange for \eq (\ref{singsp}) \cite{nad,nad1}.

\section{Numerical results}\label{numerics}

In this section we present the results of our numerical study 
of pure SU(2) gauge theory in 2+1 dimensions. 
Simulations in that theory are cheap and large Polyakov loops can be computed
without recourse to error-reducing techniques.
In order to unambiguously extract the lowest energy eigenvalues 
and their corresponding matrix elements,  
the number of timeslices $N_t$ has to be large enough for the 
ground state to dominate the exponential decay
(i.e.~to approach zero temperature, $T=1/(aN_t)\rightarrow 0$).
We work at gauge coupling $\beta=9$, which is large enough for the 
physical spectrum to be 
close to the continuum, and spatial volumes $V=24^2$ 
known to be free of finite size effects \cite{ptw,op1}.
By considering lattices with different temporal extent 
$N_t=24,26,28,30,32$, we are able to 
monitor whether the correlators are indeed described by 
single exponential decay, and cleanly separate exponents and matrix elements. 

\begin{figure}[th]
\vspace*{0.7cm}
\leavevmode
\includegraphics[width=7cm]{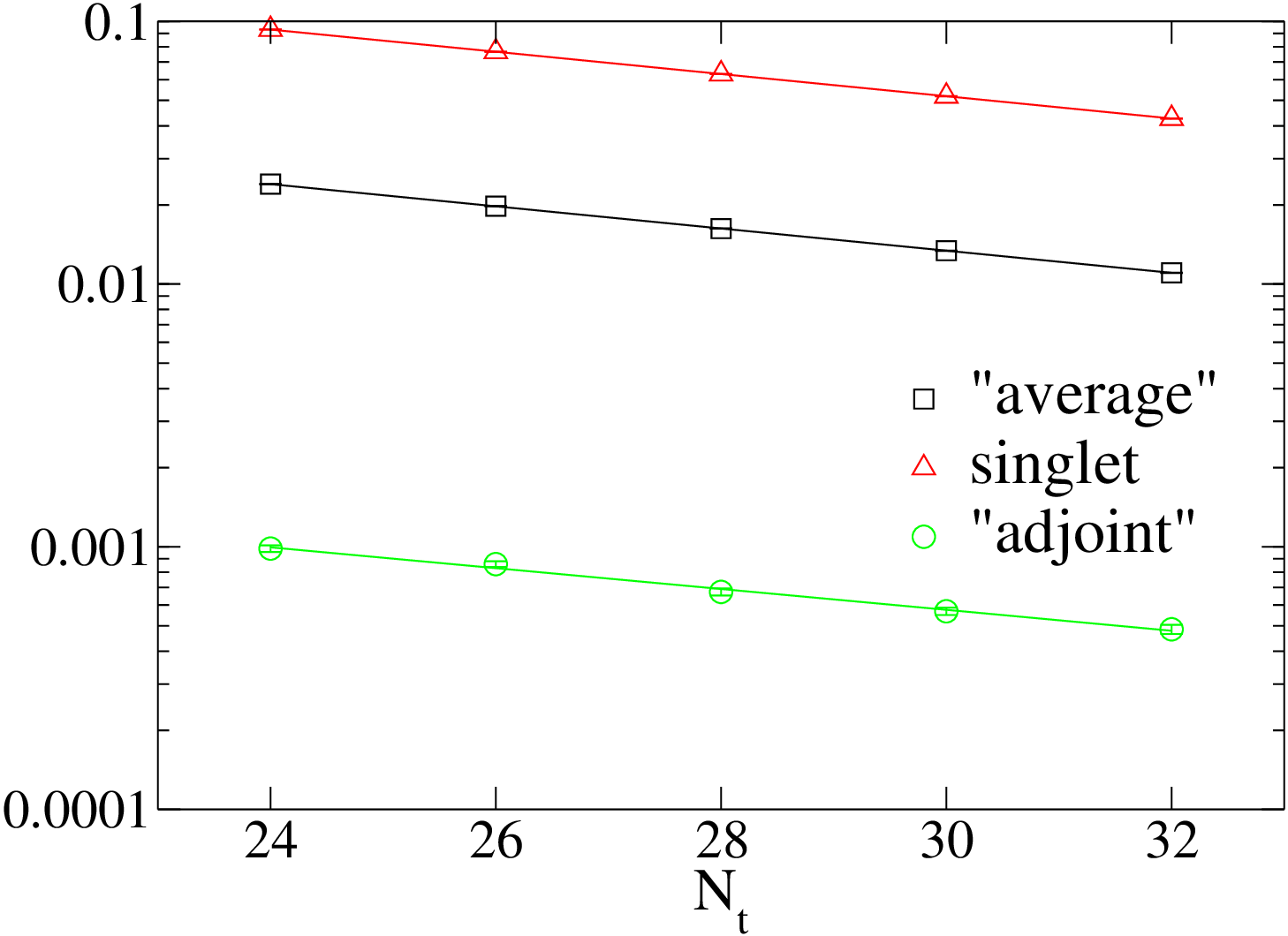}
\hspace*{1cm}
\includegraphics[width=7cm]{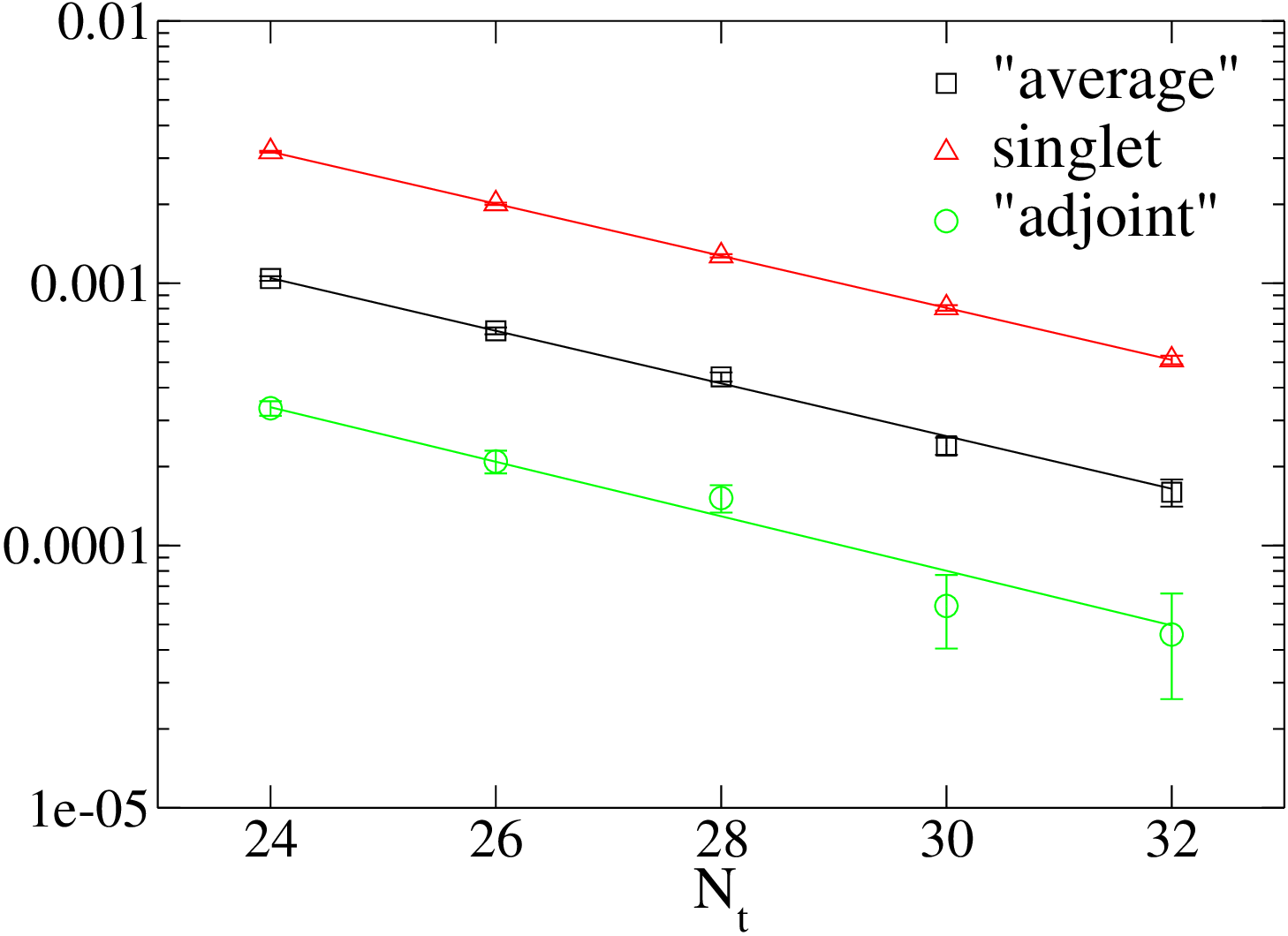}
\caption[a]{{\em
The three different correlation functions \eqs (\ref{vav},\ref{potdef}), 
constructed from
dressed Polyakov lines, for $r=a$ (left) and $r=4a$ (right);
the lines represent single exponential fits. 
}}
\la{corrs}
 \end{figure}

\begin{table}[t]
\begin{center}
\begin{tabular}{|c|*{7}{r@{.}l|}l}
\hline
\hline
$r/a$ &
\multicolumn{2}{|c|}{``average''} &
\multicolumn{2}{c|}{$\chi^2/{\rm dof}$} &
\multicolumn{2}{c|}{singlet} &
\multicolumn{2}{c|}{$\chi^2/{\rm dof}$} &
\multicolumn{2}{c|}{``adjoint''} &
\multicolumn{2}{c|}{$\chi^2/{\rm dof}$} &
\multicolumn{2}{c|}{WL \cite{op1}}\\
\hline
1 &  0&09731(2) & 0&61 & 0&09749(1) & 0&26 & 0&0925(51) & 0&94 & 0&0976(3) \\
2 &  0&1518(7)   & 0&51 & 0&1526(2)    & 0&42 & 0&144(6)      & 0&56 & 0&1529(3) \\
3 &  0&1923(18) & 0&43 & 0&1934(11)  & 1&07 & 0&1865(72) & 0&32 & 0&1935(4) \\
4 &  0&2312(83) & 1&13 & 0&2286(33)  & 1&42  & 0&239(25)   & 0&98 & 0&2280(4) \\
\hline
\hline
\end{tabular}
\end{center}
\caption[a]{{\em Fitted static potentials $a E_1(r)$ from ``average'', singlet 
and ``adjoint'' correlation functions, as well as the Wilson loop.  The values agree within errors.}}
\la{fit}
\end{table}

The result of this procedure for $r/a=1,4$ is shown in \fig \ref{corrs}, 
where all three correlation functions are plotted against $N_t$. 
Table \ref{fit} shows the resulting slope parameters, the static potentials 
$E_1(r)$, for $r/a=1,2,3,4$.
One finds that
the potential is the same for all three, and that it corresponds precisely to the
zero temperature singlet potential one obtains from a Wilson loop calculation 
at the same lattice spacing \cite{op1}. 

The matrix elements of the ground states, cf.~\eq (\ref{matel}), are
shown in \fig \ref{mat}. While our signal gets quickly noisy with growing 
distance, one can clearly discern that in the 
``average'' channel there is only the trivial constant $1/N^2$, in accord with 
our spectral decomposition.
In the singlet channel on the other hand, 
there is a non-trivial matrix element deviating significantly from one 
with growing separation, and similarly for the ``adjoint'' channel.

\begin{figure}[t]
\vspace*{1.0cm}
\centerline{\includegraphics[width=9cm]{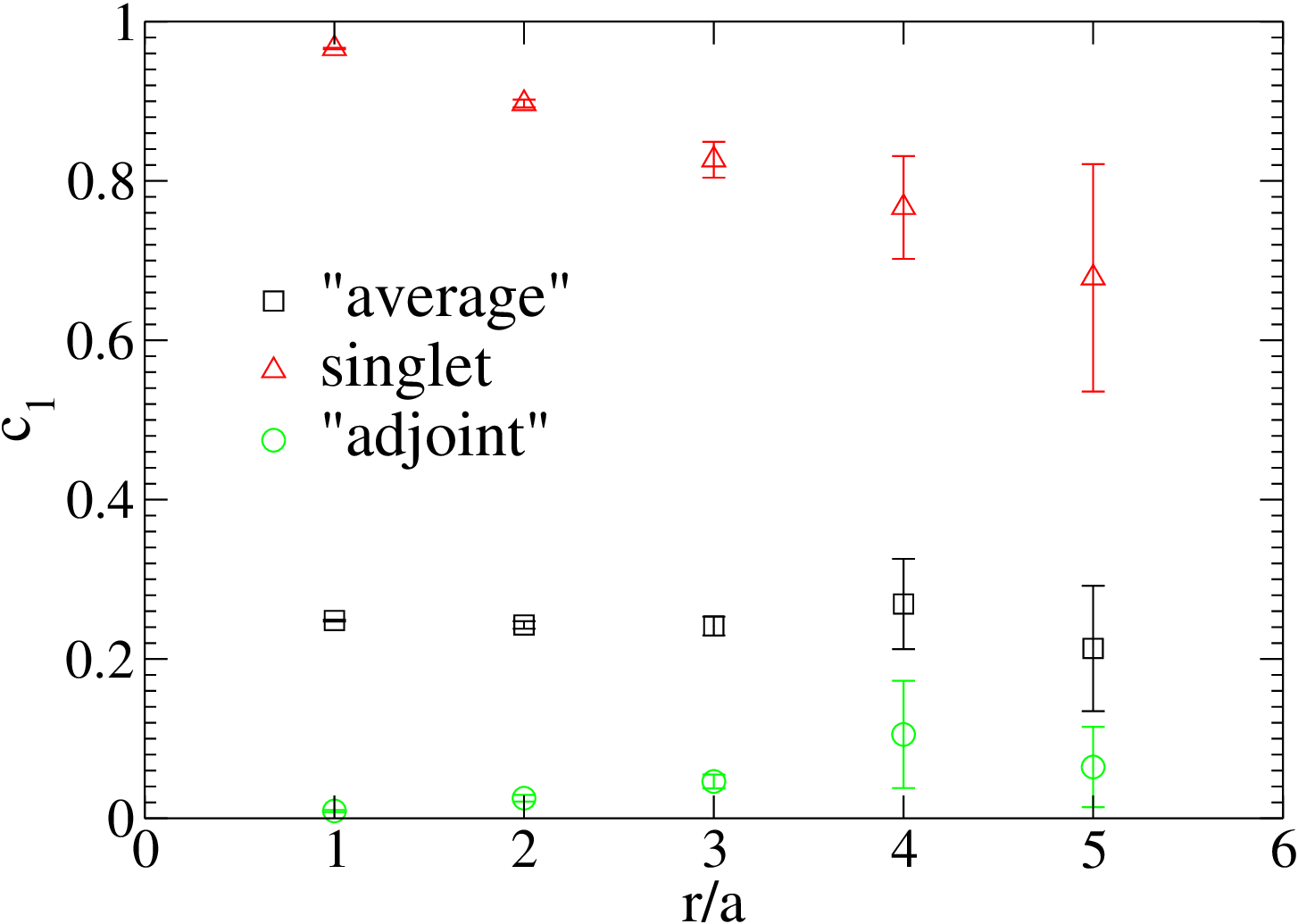}}
\caption[a]{{\em
The matrix elements of the ground state in all correlators, \eq (\ref{matel}).
}}
\la{mat}
 \end{figure}

In the light of this, how can the previous numerical results in the literature be understood?
If we just extract the potentials according to the 
prescription of formulae \eqs 
(\ref{vav},\ref{potdef}), we obtain the plot shown in \fig \ref{comp}.  
This indeed reproduces previous results which were suggestive of
three different potentials. 
However, from our analysis it is quite clear 
that this structure is due to 
the matrix elements, which get exponentiated when formulae 
\eqs (\ref{vav},\ref{potdef}) are used. 
In order to verify this claim, 
let us truncate the spectral decompositions of the
potentials to the ground state (since we are near the infinite $N_t$, or $T=0$ limit),
\ba
\ex^{-V_{\rmav}(r)/T} &=& c^{\rmav}_1 \, \ex^{-E_1(r)/T} , \\
\ex^{-V_1(r)/T} &=& c_1^1(r) \, \ex^{-E_1(r)/T} , \\
\ex^{-V_{\rmA}(r)/T} &=& c^{\rmA}_1(r) \, \ex^{-E_1(r)/T} , \label{Vadtrunc}
\ea 
with the same energy $E_1(r)$ for all channels, as predicted.
Expression (\ref{avdec}) for the decomposition of the ``average'' potential
and relation (\ref{vav}) then imply
\be\label{cad}
c^{\rmav}_1=\frac{1}{N^2},\qquad c^{\rmA}_1(r) = \frac{1-c_1^1(r)}{N^2-1} .
\ee
Thus, the would-be adjoint and average channel potentials are really
given in terms of the singlet static potential $E_1(r)$
and its matrix element
\be 
c^1_1(r)=\frac{1}{N^2}
|\langle1_{\delta\gamma}|\hu_{\gamma\delta}(\bfx,\bfy)\hu^{\dag}_{\alpha\beta}(\bfx,\bfy)
|1_{\beta\alpha}\rangle |^2  .
\ee
We have inserted the central values of our data for 
$E_1(r),c^1_1(r)$ in these formulae, \fig \ref{comp}, which indeed
reproduce the curves extracted by means of the old definition.

\begin{figure}[th]
\vspace*{1.0cm}
\leavevmode
\centerline{\includegraphics[width=9cm]{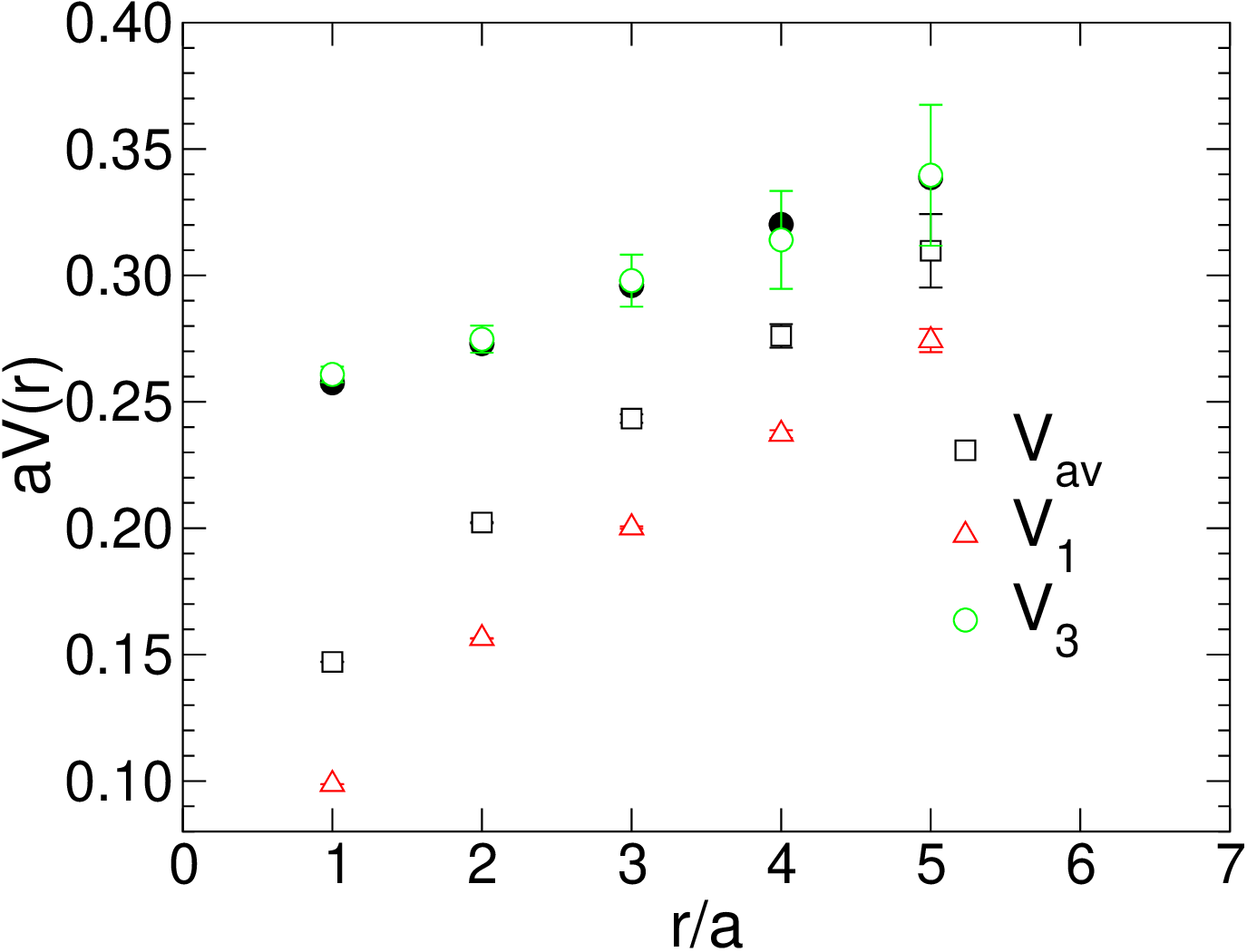}}
\caption[a]{{\em
    The potentials computed according to \eq (\ref{potdef}). The
    ``adjoint channel'' represents a combination of the singlet
    potential and its matrix element according to \eqs
    (\ref{Vadtrunc}) and (\ref{cad}), which is shown by the full
    circles.  }} \la{comp}
 \end{figure}

\section{Conclusions}

By using projection operators on Hilbert space sectors as well as a
numerical analysis of the zero temperature limit, we have shown that
Polyakov loop correlators in each of the ``average'', singlet and
``adjoint'' channels defined in the literature receive 
contributions {\it exclusively} from singlet states. Their exponential
decay in the temporal direction is thus determined by the singlet
potential and its excitations only. As expected, in the zero
temperature limit the ground state singlet potential is identical with
the one extracted from Wilson loops. Correspondingly, at finite
temperature one may extract only a singlet free energy.  At fixed
lattice spacing, its correct definition as a Boltzmann weighted sum of
exponentials without matrix elements follows from \eq (\ref{avdec}) to
be
\be
\ex^{-F_1(r)/T}=\sum_n \ex^{-E_n/T}=\langle \tr L^\dag(\bfx) \tr L(\bfy)\rangle .
\ee
Any other definition, as well as the other channels considered in the literature, exponentiate
(operator dependent) matrix elements, thus faking an $r$ and/or $T$-dependence 
which is not shared by the physical states.  

Finally, let us note that we do not dispute the existence of an
adjoint sector, $\rmF(\bfx)\otimes \bar{\rmF}(\bfy)\otimes
\rmA(\bfx_0)$, in the Hilbert space.  However, it is not probed by any
combination of Polyakov loops, but requires a different operator,
which must not close through the time boundary in order to truly
couple to adjoint states.  For a non-perturbative investigation, the
only way of saturating an open adjoint index is then by a
corresponding adjoint source.  However, in this case the resulting
object is a gluelump formed by an adjoint source coupled to two
fundamental ones in the adjoint representation, which is no longer the
situation of interest (and considered by perturbation theory), with
only two fundamental sources in the adjoint. It appears that more work
is needed to resolve these questions.

\paragraph{Note added.}

In a somewhat different use of language, it has been suggested to
interpret hybrid potentials (i.e. angular momentum excited states with
two fundamental sources \cite{MK}) as octet potentials, since they
appear to approach the gluelump spectrum in the short distance limit
\cite{lump}. According to their transformation behavior, all these
belong to the singlet sector, since the total system of static quark,
antiquark and glue is singlet under gauge transformations.  While
relevant for hybrid mesons, these potentials are therefore not useful
for describing intermediate color-charged states as they appear in
some nonrelativistic QCD computations \cite{CL}.  These latter states
have to carry an extra adjoint charge.  We thank G. Bali for
correspondence on this point.


\appendix

\section{Projectors on charge sectors} \label{projectors}

The transfer matrix $\hat T_0$ acts on the full Hilbert space of all square 
integrable wave functions, including ones transforming nontrivially under 
gauge transformations. This Hilbert space splits into 
orthogonal sectors
with charges in arbitrary representations at any lattice point.  These
are characterized by their transformation properties under the local
gauge group and can be obtained by acting with appropriate projection
operators.  The projector on gauge-invariant states (with no charges)
can be written as
\be
\hat P = \int D g\, \hat R[g],
\ee
where $\hat R[g]$ performs a gauge transformation with gauge function $g$,
$\hat R[g]\psi[U]=\psi[U^g]$. Specifically, a string between fundamental
charges at $\bfx,\bfy$ transforms as
\be\label{transformation}
\hat R[g] \hat U_{\alpha\beta}(\bfx,\bfy) \hat R^\dagger[g]
= g_{\alpha\gamma}(\bfx) \hat U_{\gamma\delta}(\bfx,\bfy) 
g^\dagger_{\delta\beta}(\bfy) \;.
\ee
The projector on states with a fundamental charge at $\bfx$ and an
anti-fundamental one at $\bfy$ is thus given by
\be
\hat P^{\rmF\otimes\bar{\rmF}} 
= N^2 \int D g\, \tr g^\dagger(\bfx)\, \tr g(\bfy) \,\hat R[g]\;.
\ee

The operator $\hp_{\alpha\beta\mu\nu}^{\rmF\otimes\bar{\rmF}}$ defined in
\eq (\ref{P33}) maps the component $\psi_{\beta\mu}$ of the
representation $\rmF\otimes\bar{\rmF}$ to the component $\psi_{\alpha\nu}$
and annihilates all other components and charge sectors.  The
components are here defined by the transformation property
\be
\hat R[g]\, \psi_{\beta\mu} = \psi_{\gamma\nu} \, 
g_{\beta\gamma}(\bfx) g^\dagger_{\nu\mu}(\bfy) \;.
\ee
A direct computation shows
\be
\hp_{\alpha\beta\mu\nu}^{\rmF\otimes\bar{\rmF}} \, \psi_{\gamma\delta}
= \frac{1}{N^2} \delta_{\beta\gamma} \delta_{\mu\delta} \,
\psi_{\alpha\nu}
\ee
and that $\hp_{\alpha\beta\mu\nu}^{\rmF\otimes\bar{\rmF}}$ annihilates all 
other charge sectors:
\be
\hp_{\alpha\beta\mu\nu}^{\rmF\otimes\bar{\rmF}} \, (1 - \hp^{\rmF\otimes\bar{\rmF}})
= 0 \;.
\ee
The normalization of $\hp_{\alpha\beta\mu\nu}^{\rmF\otimes\bar{\rmF}}$ has
conveniently been chosen such that
\be
\sum_{\beta\mu} \hp_{\alpha\beta\mu\nu}^{\rmF\otimes\bar{\rmF}} \, 
\psi_{\beta\mu}
= \psi_{\alpha\nu} \;.
\ee
The projection operators
$N^2\hp_{\alpha\alpha\mu\mu}^{\rmF\otimes\bar{\rmF}}$ (no sum) provide a
decomposition of $\hp^{\rmF\otimes\bar{\rmF}}$,
\be
\hat P^{\rmF\otimes\bar{\rmF}} 
= N^2 \sum_{\alpha\mu} \hp_{\alpha\alpha\mu\mu}^{\rmF\otimes\bar{\rmF}} \;.
\ee

\end{document}